\documentclass[final]{svjour2}
\usepackage{graphicx}
\usepackage{rotating}
\usepackage{amssymb}
\usepackage{mathptmx}
\usepackage[numbers]{natbib}
\makeatletter
\journalname{Journal of Low Temperature Physics}

\usepackage{color}

\bibpunct{}{}{,}{s}{}{,}

\begin{document}

\newcommand{\hdblarrow}{H\makebox[0.9ex][l]{$\downdownarrows$}-}
\title{Characterisation of a TES-based X-ray microcalorimeter in the energy
range from 150 to 1800 eV using an Adiabatic Demagnetisation
Refrigerator}

\author{L. Gottardi \and Y. Takei  \and J. van der Kuur \and P.A.J. de
  Korte \and H.F.C.Hoevers \and D. Boersma \and  M. Bruijn 
  \and W. Mels \and M. L. Ridder \and D. Takken \and H. van Weers}

\institute{SRON Netherlands Institute for Space Research,\\
  Sorbonnelaan 2, 3584CA, Utrecht, The Netherlands\\ Tel.:+31-30-2538561\\ Fax:+31-30-2540860\\
\email{l.gottardi@sron.nl}}
                
\date{23.07.2007}

\maketitle

\keywords{TES, X-ray calorimeter, superconductivity, synchrotron radiation}

\begin{abstract}

We characterised  a TES-based  X-ray microcalorimeter in  an adiabatic
demagnetisation  refrigerator (ADR)  using synchrotron  radiation. The
detector  response and  energy  resolution was  measured at  the
beam-line in the PTB radiometry laboratory at the electron storage ring
BESSY II in the range from 200  to 1800 eV. We present and discuss the
results of the energy resolution measurements as a function
of  energy, beam intensity  and detector  working point.  The measured
energy resolution  ranges between  1.5 to 2.1  eV in  the investigated
energy  range and is  weakly dependent  on the  detector set  point. A
first analysis shows a count-rate capability, without considerable
 loss of performance, of about 500 counts per second. 

PACS numbers:07.20.Fw,07.85.Nc,85.25.Am,85.25.Oj,95.55.-n 
\end{abstract}

\section{Introduction}

We are developing an imaging array of
 Transition Edge Sensor (TES) microcalorimeters
for future X-ray astronomy mission like EDGE (Explorer of the
Diffuse emission and Gamma-ray burst Explosions)\cite{EDGE}, XEUS
\cite{XEUS} and Constellation-X\cite{CosX}. 
The experiment described here is part of the European-Japanese project
EURECA, which aims to demonstrate technological readiness of a 5 x 5
pixel array of TES-based micro-calorimeters read-out by two
SQUID-amplifier channels using frequency-domain-multiplexing (FDM) \cite{PietLTD12}. 
We discuss here the details of the detector set-up and the first
results of the tests made using  synchrotron radiation at the PTB beamline of the BESSY II \cite{PTBBESSY}. 
 The  aim  of the campaign was to test the sensor behaviour  to
 investigate crucial detector issues like calibration of the energy
 scale, non linearity, large signal analysis  and energy resolution
 retrieval from pile-up events.

\section{Experimental details}

\begin{figure}
\begin{center}
\includegraphics[%
width=1\linewidth,
keepaspectratio]{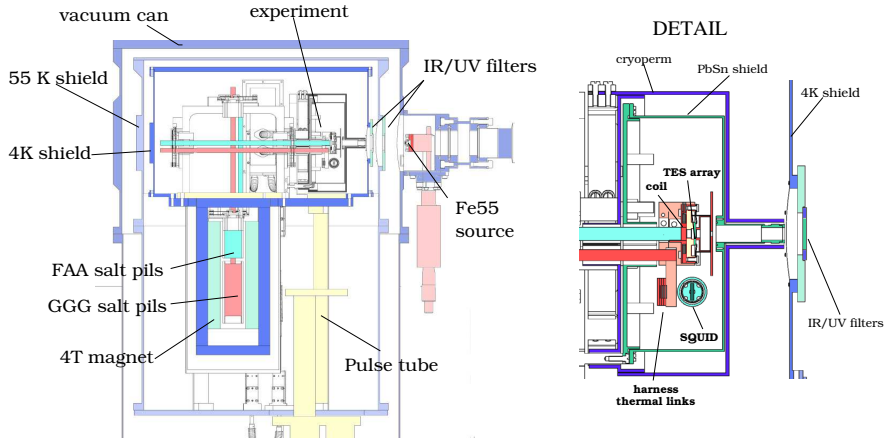}
\end{center} 
\caption{(Color online) Cross section of the detector integrated in a
  two-stage Adiabatic Demagnetisation Refrigerator precooled by a Pulse Tube    }
\label{set-up}
\end{figure}

A cross section of the experimental set-up is shown in Fig.~\ref{set-up}.
Tab.~\ref{setuptable} gives a summary of the cooler and sensor main parameters

  \begin{table}[htbp]
  \begin{center} \small
    \begin{tabular}{c c c c}  
\hline  \multicolumn{2}{c}{ADR} 
& \multicolumn{2}{c}{TES sensor}  \vspace{0.05 cm}\\ 
{\it parameter} & {\it value} & {\it parameter} & {\it value}
\vspace{0.05 cm}
\\ \hline \vspace{0.2 cm} Pulse tube Temperature &  $ 3.5$ K & Ti/Au area  & $146\mu
m \times 150\mu m$ 
\\  PT cooling power &  $350$ mW &  Ti/Au thickness &
$25$ nm (Ti),$\;$ $50$ nm (Au) \\  GGG salt pills
T& 880 mK & Cu abs. area&
$100\mu m\times 100\mu m$  \\  FAA salt pills T& $48$ mK  & Cu
abs. thickness & $1000$nm  
\\  hold time  at $73\, mK$ & $\sim 24$ h  & $T_C$& 100 mK \\ magnetic field & 4 T& $R_N$
 & $143$ m$\Omega$ \\&& G at $T_C$ & $0.36$ nW/K \\  && C
& 0.3 pJ/K\\  && Power at $T_{bath} =73\,mK$& 6.5 pW\\
\hline
     \end{tabular}
     \caption{ADR and TES microcalorimeter main parameters. At a TES
     resistance of $R\simeq 0.5 R_N$ the thermometer sensitivity  
     to temperature and current is respectively  $\alpha\simeq 380 $ and $\beta\sim 2$,[\cite{YohLTD12}\cite{HenkLTD12}]}
     \label{setuptable}
  \end{center}
\end{table}
The sensor is  a single-pixel of a 5x5 array and consists of a Cu
absorber on top of a $Ti/Au$ TES deposited on a $Si_xN_y$ membrane,
which  provides a weak thermal link to the ADR bath temperature. 
The TES is voltage-biased with a $R_{th} = 10\, m\Omega$ (thevenin
equivalent) load resistance, and kept at the transition by negative
electro-thermal feedback (ETF)  \cite{Irwin}.    

The current through the TES was
measured by a 100-SQUID array  \footnote{kindly provided by NIST} operated in flux locked
loop (FLL). The inductance of the SQUID input coil is $L_s\sim 70$ nH.
The SQUID array is directly read out by a low noise commercial
PTB-Magnicon electronics \footnote{Magnicon-PTB, www.magnicon.com}.

The detector has been
integrated in a Janis two-stage ADR cooler
precooled at 3.5 K by a mechanical Cryomech Pulse Tube (PT).

To reduce the electro-magnetic interferences from the external
environment, the detector cold-head, the harness and the room
temperature electronics is fully enclosed in a dedicated
Faraday cage. 
In the cold stage, the Faraday cage is formed by a double
shield anchored at $3.5 K$ consisting of a cryoperm shield and a
copper load-tin plated superconducting shield. 
The cryostat is foreseen of windows along the optical path of the
detector. We employ two 150 nm  aluminium filters, each supported on a
200 nm parylene substrate, respectively placed on the 60 and 3.5 K cryostat shields.

We characterised the detector using the soft X-ray two plane grating
monochromator beam-line (SX700) in the PTB laboratory at BESSY II
synchrotron facility in
Berlin \cite{PTBBESSY}. The photons energy can be selected with high 
resolution ($< 0.5$ eV) in the range from 50 to 1800 eV. The radiation flux can be 
easily change from a few tenths to thousands of photons per second.

\section{Results}

We report here the results of  the  measurement campaign at PTB-BESSY.
A detailed characterisation of the single pixel thermal and electrical
responsivity can be found in
\cite{YohLTD12} \cite{HenkLTD12}. 

We study the microcalorimeter detector by measuring  the energy
 resolution in the energy range from $150$ eV to $1.8$ keV 
 and for the microcalorimeter optimum bias point observed at
 $R/R_N=0.46$ and zero perpendicular magnetic field,
$B_\bot \sim 0$ mG obtained by cancelling the remanent magnetic
 field with an auxiliary superconducting coil. The base temperature was $T_{bath}=73$ mK  and
stabilised at a level of $7 \mu K$ rms for
 all the measurements described in this paper.
The dependence  from the TES bias point and the applied perpendicular
static magnetic field was also studied and  will be discussed.

\begin{figure}
\begin{center}
\includegraphics[%
  width=1\linewidth,
  keepaspectratio]{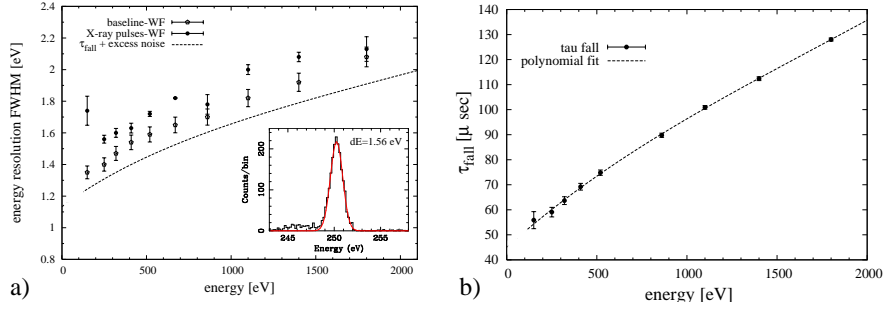}
\end{center}
\caption{{\bf  a)}  X-ray (full  circle)  and  baseline (open  circle)
  energy resolution  as a  function of the  photon energy for  the TES
  optimal working point. The photon  flux was of about 50 photons/sec.
  The  theoretical (M=0) energy  resolution for  this detector  at the
  working point discussed here,  is estimated to be $\Delta E_{th}\sim
  0.6$ eV. {\bf b)} Pulse fall time $\tau_{fall}$ as a function of
energy. The dashed curve is the polynomial curve fitting the data.}
\label{dEvsE}
\end{figure}

In  Fig.~\ref{dEvsE}.{\bf a.} the energy resolution is
shown as a function of the incident photon energy.
In the graph we plotted with full circles the X-ray energy resolution
obtained after applying the matched filter to each single photon. The open
circles show the baseline resolution obtained by filtering the
detector noise with the same filter used for the X-ray pulses. When
the instrumental resolution of the detector is dominated by random
noise, the baseline resolution is equal to the X-ray resolution.
At 250 eV we measured the best energy resolution of $1.52\pm 0.03$ eV
(see insert of Fig.~\ref{dEvsE}.{\bf a.}).
 In the presence  of electro-thermal feedback (ETF) and  for large ETF
loop gain ${\cal L}_0$, the  energy resolution can be approximated, in
the small signal analysis, as
\begin{equation}\label{dEfwhm}
\Delta            E_{FWHM}\simeq           2.355\sqrt{4kT_c\tau_{fall}
  P_{J_0}}(n(1+M^2)/2)^{1/4},
\end{equation}
where  $T_C$ is  the TES  transition temperature,$\alpha$  and $\beta$
describe  respectively  the  temperature  and  current  dependent  TES
transition steepness  and $P_{J_0}=I^2_0 R$ is  the steady-state Joule
power.  The  pulse  fall   time  is  calculated  to  be  $\tau_{fall}=
\tau_0/(1+{\cal  L}_0/(1+\beta))$, where  $\tau_0=C/G$ and  $n=3.7$ is
determined by the heat transport  mechanism. M represents the ratio of
the level of  excess noise to the Johnson noise  as described by Ullom
et  al. \cite{Ullom}.  The theoretical  energy resolution  ($M=0$) for
this sensor is estimated to be $\Delta E_{th}\sim 0.6$ eV.

The observed dependency of the energy resolution on the photons energy
is  related to the  change of  the pulse  fall time  as a  function of
energy shown  in Fig.~\ref{dEvsE}.{\bf b}. In a  neighbourhood of this
particular bias  point the  sensor parameters like  $\alpha$, $\beta$,
and ${\cal  L}_0$ are changing quite  rapidly \cite{YohLTD12}, leading
to a  strong dependence of  the $\tau_{fall}$ on the  signal amplitude
and thus  the energy.  From Eq.~\ref{dEfwhm} and  the polynomial curve
obtained to  fit the experimental  measurements of $\tau_{fall}$  as a
function   of   energy  we   derive   the   dashed   curve  shown   in
Fig.~\ref{dEvsE},  which  gives  the  expected  energy  resolution  in
presence of excess noise for our sensor.

We used $M=3$, $P_{J_0}=6.5$ pW and we assumed them to be constant for
all the energies.  The behaviour  of the X-ray and baseline resolution
is  qualitatively  explained   by  Eq.~\ref{dEfwhm}.  The  discrepancy
between the measured baseline resolution and the calculated resolution
is of only about $15 \%$. This could be due to the thermal fluctuation
noise (not  include in  the estimated  $dE$ due to  the presence  of a
dangling    heat   capacitance    thermally    decoupled   from    the
TES\cite{YohLTD12}.

The X-ray energy resolution is about $20\%$ worse than the
baseline resolution. This may indicate that the applied matched filter
is not optimal due to the non-stationary property of the noise. In
that case one
would expect however the difference to be larger at higher energy.
 Drifts due to fluctuations of the bath temperature or magnetic field could
also explain the discrepancy.

\begin{figure}
\begin{center}
\includegraphics[%
  width=0.65\linewidth,
  keepaspectratio]{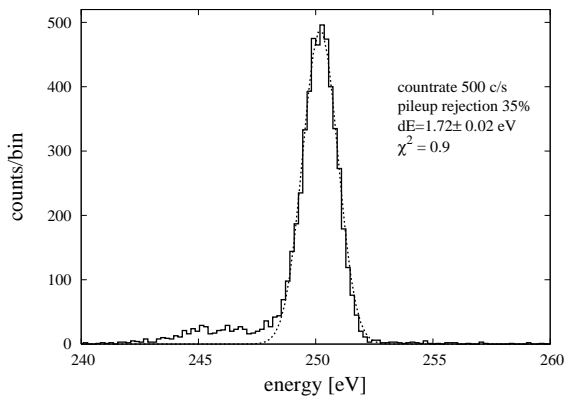}
\end{center}
\caption{Energy spectra for the optimal bias point and 
  photon flux of about 500 photons/sec. We rejected about $30\%$ of
  events due to pile-up}
\label{highflux}
\end{figure}

Fig.~\ref{highflux} shows the energy distribution at $250$ eV for
 a count-rate as large as 500 photons per second. When large photon fluxes are applied to the calorimeter
 a slightly worse energy resolution is observed. However, as shown in
 the figure, for a radiation flux as large as 500 photons/sec we could still
measure an energy resolution of  $1.78\pm 0.02$ eV. To produce the
spectra we rejected about $30\%$ of events due to pile-up. The energy
 degradation could be due to a non perfect pile-up rejection or non-linear effects
in the detector caused by the large number of photons reaching the
 microcalorimeter. The excess energy counts observed
in the low energy side of the spectrum are due to photons not directly
absorbed in the microcalorimeter absorber. This is consequence of the
fact that the collimator opening is slightly larger than the absorber size.
The data analysis for high fluxes is still in progress. The complete
results will be presented in a future paper.

\begin{figure}
\begin{center}
\includegraphics[%
  width=0.65\linewidth,
  keepaspectratio]{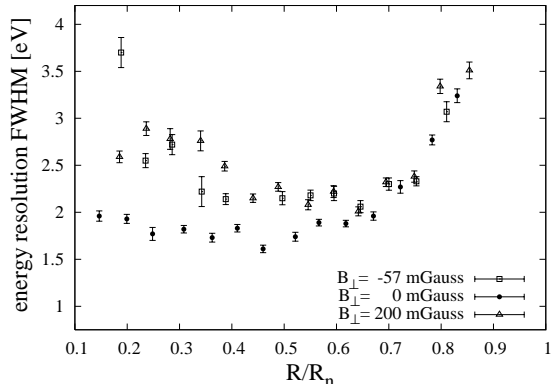}
\end{center}
\caption{ X-ray  energy resolution for photons of energy $E=250eV$ as
  a function of TES dynamic resistance R and the perpendicular magnetic
  field.}
\label{dEvsbias}
\end{figure}

The dependency of the photon energy resolution on the TES bias point
and magnetic field is shown in Fig.~\ref{dEvsbias} for an energy of
$250$ eV. The best energy resolution is observed for $R/R_N\simeq
0.46$ at zero perpendicular magnetic field $B_\bot$ in the TES.
We observed a degradation of the energy resolution of about $20\%$
from the optimum when the
TES is biased between $20\%$ and $60\%$ of $R_N$. This is consistent
with the results reported by Y.Takei et al. \cite{YohLTD12} on this
sensor, where it is shown that the responsivity scales accordingly
with the noise in this bias region.

   When a magnetic filed
is applied the energy resolution becomes worse due to a lower
thermometer superconducting transition steepness $\alpha$. For a
magnetic field as large as $B_\bot =\pm 200$ mG the energy
resolution becomes worse of about $50\%$.


\section{Conclusions}
We presented the response and energy resolution measurements of a single pixel TES
microcalorimeter operating in an ADR performed at the
beam-line in the PTB radiometry laboratory at the electron storage ring
BESSY II in the range from 200 to 1800 eV. 
At $E=250$ eV we measured an energy resolution of $1.52 \pm 0.03$
eV at the best bias point corresponding to a TES resistance
$R/R_N=0.46$, for zero perpendicular magnetic field and for a count
rate of 50 photons/sec. The energy resolution was found to be dependent on the photon
energy and is related to the change of the pulses fall
time. A small degradation of the energy resolution was observed for
large count-rate. With 500 photons/sec reaching the detector an energy
resolution of $1.78\pm 0.02$ eV at $250$ eV was measured with a
pile-up rejection of $30\%$.
Further we found a weak dependency of the energy resolution on
the detector working point consistent with the responsivity and noise
measurement reported in Ref.~\cite{YohLTD12}.
The data analysis is still in progress. We aim to fully investigate crucial detector issues like the
calibration of the energy scale, non linearity, energy resolution
retrieval from pile-up events and large signal analysis.  

\begin{acknowledgements}
We are grateful to Marcos Bavdaz and Didier Martin from ESA for giving us the possibility to use the beam-line at PTB-BESSY and to people
from the PTB radiometry laboratory for their help during the
measurements. We thank Paul Lowes and
Arjan de Klein for their precious technical help and Chris Whitford
for providing the IR/UV-filters.
This work is financially supported by the Dutch Organization for Scientific Research (NWO).
\end{acknowledgements}


\end{document}